\documentclass[prl,preprint,superscriptaddress,showpacs,amssymb,amsmath,amsfonts,aps]{revtex4}
\usepackage{graphicx}
\usepackage{dcolumn}
\usepackage{bm}

\begin{document}

\title{Proton source size measurements in the $eA \rightarrow e'ppX$
reaction}

\newcommand*{\ITEP }{ Institute of Theoretical and Experimental
 Physics, Moscow, 117218, Russia} 
\affiliation{\ITEP } 

\newcommand*{\CZA }{ Institute of Physics, Czech Academy of Sciences,
 Na Slovance 2, 18040 Prague 8, Czech Republic} 
\affiliation{\CZA } 

\newcommand*{\ASU }{ Arizona State University, Tempe, Arizona 
85287-1504, USA} 
\affiliation{\ASU } 

\newcommand*{\SACLAY }{ CEA-Saclay, Service de Physique Nucl\'eaire,
 F91191 Gif-sur-Yvette, Cedex, France} 
\affiliation{\SACLAY } 

\newcommand*{\UCLA }{ University of California at Los Angeles,
 Los Angeles, California  90095-1547, USA} 
\affiliation{\UCLA } 

\newcommand*{\CMU }{ Carnegie Mellon University,
 Pittsburgh, Pennsylvania 15213, USA} 
\affiliation{\CMU } 

\newcommand*{\CUA }{ Catholic University of America,
 Washington, D.C. 20064, USA} 
\affiliation{\CUA } 

\newcommand*{\CNU }{ Christopher Newport University,
 Newport News, Virginia 23606, USA} 
\affiliation{\CNU }  

\newcommand*{\UCONN }{ University of Connecticut, Storrs, 
Connecticut 06269, USA} 
\affiliation{\UCONN } 

\newcommand*{\DUKE }{ Duke University, Durham, North
 Carolina 27708-0305, USA} 
\affiliation{\DUKE } 

\newcommand*{\ECOSSEE }{ Edinburgh University, Edinburgh EH9 3JZ, United Kingdom} 
\affiliation{\ECOSSEE } 

\newcommand*{\FIU }{ Florida International University, Miami,
 Florida 33199, USA} 
\affiliation{\FIU } 

\newcommand*{\FSU }{ Florida State University, Tallahassee, 
Florida 32306, USA} 
\affiliation{\FSU } 

\newcommand*{\GEISSEN }{ Physikalisches Institut der 
Universitaet Giessen, 35392 Giessen, Germany} 
\affiliation{\GEISSEN } 

\newcommand*{\GWU }{ The George Washington University,
 Washington, DC 20052, USA} 
\affiliation{\GWU }

\newcommand*{\ECOSSEG }{ University of Glasgow, Glasgow G12 8QQ, 
United Kingdom} 
\affiliation{\ECOSSEG } 

\newcommand*{\INFNFR }{ INFN, Laboratori Nazionali
 di Frascati, Frascati, Italy} 
\affiliation{\INFNFR } 

\newcommand*{\INFNGE }{ INFN, Sezione di Genova,
 16146 Genova, Italy} 
\affiliation{\INFNGE } 

\newcommand*{\ISU }{ Idaho State University, Pocatello, 
Idaho 83209, USA} 
\affiliation{\ISU } 

\newcommand*{\ORSAY }{ Institut de Physique Nucleaire
 ORSAY, Orsay, France} 
\affiliation{\ORSAY } 

\newcommand*{\JMU }{ James Madison University, Harrisonburg,
 Virginia 22807, USA} 
\affiliation{\JMU } 

\newcommand*{\KYUNGPOOK }{ Kungpook National University,
 Taegu 702-701, South Korea} 
\affiliation{\KYUNGPOOK } 

\newcommand*{\MIT }{ Massachusetts Institute of Technology,
 Cambridge, Massachusetts  02139-4307, USA} 
\affiliation{\MIT } 

\newcommand*{\UMASS }{ University of Massachusetts, Amherst,
 Massachusetts  01003, USA} 
\affiliation{\UMASS } 

\newcommand*{\MOSCOW }{ Moscow State University, General Nuclear 
Physics Institute, 119899 Moscow, Russia} 
\affiliation{\MOSCOW } 

\newcommand*{\UNH }{ University of New Hampshire, Durham,
 New Hampshire 03824-3568, USA} 
\affiliation{\UNH } 

\newcommand*{\NSU }{ Norfolk State University, Norfolk,
 Virginia 23504, USA} 
\affiliation{\NSU } 

\newcommand*{\OHIOU }{ Ohio University, Athens, Ohio 
 45701, USA} 
\affiliation{\OHIOU } 

\newcommand*{\ODU }{ Old Dominion University, Norfolk,
 Virginia 23529, USA} 
\affiliation{\ODU } 

\newcommand*{\PENN }{ Penn State University, University Park,
 Pennsylvania 16802, USA} 
\affiliation{\PENN }
 
\newcommand*{\PITT }{ University of Pittsburgh, Pittsburgh,
 Pennsylvania 15260, USA} 
\affiliation{\PITT } 

\newcommand*{\ROMA }{ Universita' di ROMA III,
 00146 Roma, Italy} 
\affiliation{\ROMA } 

\newcommand*{\RPI }{ Rensselaer Polytechnic Institute,
 Troy, New York 12180-3590, USA} 
\affiliation{\RPI } 

\newcommand*{\RICE }{ Rice University, Houston,
 Texas 77005-1892, USA} 
\affiliation{\RICE } 

\newcommand*{\UTEP }{ University of Texas at El Paso, El Paso, 
Texas 79968, USA}

\newcommand*{\URICH }{ University of Richmond, Richmond,
 Virginia 23173, USA} 
\affiliation{\URICH } 

\newcommand*{\SCAROLINA }{ University of South Carolina, Columbia,
 South Carolina 29208, USA} 
\affiliation{\SCAROLINA } 

\newcommand*{\ISUP }{ Idaho State University, Pocatello,
 Idaho 83209, USA} 
\affiliation{\ISUP } 

\newcommand*{\JLAB }{ Thomas Jefferson National Accelerator
 Facility, Newport News, Virginia 23606, USA} 
\affiliation{\JLAB } 

\newcommand*{\UNIONC }{ Union College, Schenectady, NY 12308, USA} 
\affiliation{\UNIONC } 

\newcommand*{\VT }{ Virginia Polytechnic Institute and State
 University, Blacksburg, Virginia   24061-0435, USA} 
\affiliation{\VT } 

\newcommand*{\VIRGINIA }{ University of Virginia, Charlottesville,
 Virginia 22901, USA} 
\affiliation{\VIRGINIA } 

\newcommand*{\WM }{ College of Willliam and Mary, Williamsburg,
 Virginia 23187-8795, USA} 
\affiliation{\WM } 

\newcommand*{\YEREVAN }{ Yerevan Physics Institute, 375036 Yerevan,
 Armenia} 
\affiliation{\YEREVAN } 

\newcommand*{\NOWUNIONC }{ Department of Physics, Schenectady, 
NY 12308, USA}

\newcommand*{\NOWCNU }{ Christopher Newport University,
 Newport News, Virginia 23606, USA}

\newcommand*{\NOWNCATU }{ North Carolina Agricultural and Technical 
State University, Greensboro, NC 27411, USA}

\newcommand*{\NOWECOSSEG }{ University of Glasgow, 
Glasgow G12 8QQ, United Kingdom}

\newcommand*{\NOWSCAROLINA }{ University of South Carolina, 
Columbia, South Carolina 29208, USA}

\newcommand*{\NOWJLAB }{ Thomas Jefferson National Accelerator
Facility, Newport News, Virginia 23606, USA}

\newcommand*{\NOWITEP }{ Institute of Theoretical and Experimental 
Physics, Moscow, 117259, Russia}

\newcommand*{\NOWOHIOU }{ Ohio University, Athens, Ohio  45701, USA}

\newcommand*{\NOWFIU }{ Florida International University, Miami,
Florida 33199, USA}

\newcommand*{\NOWINFNFR }{ INFN, Laboratori Nazionali 
di Frascati, Frascati, Italy}

\newcommand*{\NOWCMU }{ Carnegie Mellon University, 
Pittsburgh, Pennsylvania 15213, USA}

\newcommand*{\NOWINDSTRA }{ Systems Planning and 
Analysis, Alexandria, Virginia 22311, USA}

\newcommand*{\NOWASU }{ Arizona State University, Tempe, 
Arizona 85287-1504, USA}

\newcommand*{\NOWCISCO }{ Cisco, Washington, DC 20052, USA}

\newcommand*{\NOWUK }{ University of Kentucky, LEXINGTON, 
KENTUCKY 40506, USA}

\newcommand*{\NOWSACLAY }{ CEA-Saclay, Service de Physique 
Nucl\'eaire, F91191 Gif-sur-Yvette,Cedex, France}

\newcommand*{\NOWMOSCOW }{ Moscow State University, General 
Nuclear Physics Institute, 119899 Moscow, Russia}

\newcommand*{\NOWRPI }{ Rensselaer Polytechnic Institute, 
Troy, New York 12180-3590, USA}

\newcommand*{\NOWDUKE }{ Duke University, Durham, 
North Carolina 27708-0305, USA}

\newcommand*{\NOWUNCW }{ North Carolina, USA}

\newcommand*{\NOWHAMPTON }{ Hampton University, 
Hampton, VA 23668, USA}

\newcommand*{\NOWTulane }{ Tulane University, New Orleans, 
Lousiana  70118, USA}

\newcommand*{\NOWORSAY }{ Institut de Physique 
Nucleaire ORSAY, Orsay, France}

\newcommand*{\NOWGEORGETOWN }{ Georgetown University, 
Washington, DC 20057, USA}

\newcommand*{\NOWCUA }{ Catholic University of America, 
Washington, D.C. 20064, USA}

\newcommand*{\NOWJMU }{ James Madison University, 
Harrisonburg, Virginia 22807, USA}

\newcommand*{\NOWURICH }{ University of Richmond, 
Richmond, Virginia 23173, USA}

\newcommand*{\NOWCALTECH }{ California Institute of 
Technology, Pasadena, California 91125, USA}

\newcommand*{\NOWVIRGINIA }{ University of Virginia, 
Charlottesville, Virginia 22901, USA}

\newcommand*{\NOWYEREVAN }{ Yerevan Physics Institute, 
375036 Yerevan, Armenia}

\newcommand*{\NOWRICE }{ Rice University, Houston, 
Texas 77005-1892, USA}

\newcommand*{\NOWINFNGE }{ INFN, Sezione di Genova, 
16146 Genova, Italy}

\newcommand*{\NOWROMA }{ Universita' di ROMA III, 
00146 Roma, Italy}

\newcommand*{\NOWBATES }{ MIT-Bates Linear Accelerator 
Center, Middleton, MA 01949, USA}

\newcommand*{\NOWVSU }{ Virginia State University, 
Petersburg,Virginia 23806, USA}

\newcommand*{\NOWORST }{ Oregon State University, 
Corvallis, Oregon 97331-6507, USA}

\newcommand*{\NOWGWU }{ The George Washington University, 
Washington, DC 20052, USA}

\newcommand*{\NOWMIT }{ Massachusetts Institute of Technology, 
Cambridge, Massachusetts  02139-4307, USA}

\newcommand*{\Deceased }{ Deceased}
\altaffiliation{\Deceased }

  
\author{A.V.~Stavinsky}
     \affiliation{\ITEP}
\author{K.R.~Mikhailov}
     \affiliation{\ITEP}
\author{R.~Lednicky}
     \affiliation{\CZA}
\author{A.V.~Vlassov}
     \affiliation{\ITEP}
\author{G.~Adams}
     \affiliation{\RPI}
\author{P.~Ambrozewich}
     \affiliation{\FIU}
\author{E.~Anciant}
     \affiliation{\SACLAY}
\author{M.~Anghinolfi}
     \affiliation{\INFNGE}
\author{B.~Asavapibhop}
     \affiliation{\UMASS}
\author{G.~Asryan}
     \affiliation{\YEREVAN}
% ...

\collaboration{The CLAS Collaboration}
     \noaffiliation

\date{\today}

\begin{abstract}
Two-proton correlations at small relative momentum  $q$ were 
studied in the $eA(^3$He, $^4$He, C, Fe)$\rightarrow e'ppX$ reaction 
at $E_0 = 4.46$ GeV using the CLAS detector at Jefferson Lab.
The  enhancement of  the correlation  function  at small $q$
was found to be in accordance with theoretical expectation. 
Emission region sizes were extracted and proved to be dependent on $A$ 
and proton momentum. The size of the two-proton emission region
on the lightest possible nucleus, He, was measured for the first
time.

\end{abstract}
\maketitle

One of the outstanding issues in nuclear physics
is the nature of dense and(or) hot nuclear matter~\cite{Baldin,QGP}.
There are strong experimental indications~\cite{Kiselev,BAS89} that
density fluctuations of nuclear matter manifest themselves in so called
``cumulative processes'' in which 
particles are produced in the kinematical region forbidden 
for interactions with a single motionless nucleon,
and hence more than one target nucleon must be involved.
Cumulative particle spectra remain unexplained when 
finite temperature Fermi-gas momentum distributions
are considered~\cite{Amado}, leading to the association 
of the reaction strength in this kinematic region with 
density fluctuations or correlations.
These objects can be described
in various ways~\cite{Bloha,Efrem,Burov,frankfurt,qbag},
but all authors consider them to be fluctuations. 
In this paper, we will not rely on a specific model, 
and, following Blokhintsev~\cite{Bloha}, 
Efremov~\cite{Efrem} and others,
 will, for simplicity, refer to this type of object
 as a ``flucton''. The production of an energetic nucleon 
pair from a nucleus is also an example of a cumulative 
process, and therefore can be used to study fluctons.

Pairs of nucleons can also be produced 
due to rescattering of the emitted
particles on other nucleons in the nucleus. 
Cascade calculations~\cite{Kopel} also fail to describe 
the whole set of experimental data, but rescattering 
can affect experimental spectra and particle correlations.
The relative importance of rescattering processes depends 
on the mass number $A$ of the nucleus. We believe that 
extrapolation to the smallest $A$ will provide reliable 
information on the true properties of the flucton.  

To estimate the density of the flucton, one needs to 
measure its size and the number of contributing nucleons,
the minimum number of which can be determined from the kinematics. 
The flucton size is expected to be commensurate with the size
of a nucleon~\cite{Bloha,Burov,Efrem}.

The two-particle correlations at small
relative momentum $ \vec {q} = \vec{p_1} - \vec{p_2}$
($ \vec{p_1}$ and $ \vec{p_2}$ are the individual proton
momenta in the pair reference frame) are sensitive to the source size
~\cite{KP72,K74,KOO77,LL82}(see also the reviews~\cite{rew}).
We will use the term ``femtoscopy'' (1 fm $= 10^{-15}$ m) 
for the study of source sizes within nuclei in analogy 
with microscopy.

 The two-proton correlations at small $q$  
was theoretically described in~\cite{KOO77,LL82}. 
The interference of identical 
particles~\cite{KP72,K74}, as well as Coulomb and 
strong final  state interactions (FSIs) \cite{WATSON}
were taken into account.  
Strong FSIs are dominant, causing the increasing of the
pair production cross section near $q \sim 0.04$ GeV/c. 
The intensity of the effect depends inversely
on the root mean square radius 
$r_{\rm RMS}$ of the source from which the protons
are emitted.

It should be noted that here we understand the FSI 
as the interaction in the two-proton system 
at small relative momentum only. The interaction time
in this system is much larger than the characteristic
collision time and so this system can be considered
in isolation of other particles and described by
the same wave function as in the scattering problem
(up to the opposite direction of the relative momentum vector). 
For proton interactions with 
other particles during the  collision process
we  use another term - rescattering. Rescatterings are
generally characterized by much higher momentum transfers
and correspondingly shorter time scales. They are essentially
localized and can be considered as new emission points.
FSIs are our ``tool'' to measure the flucton size
while the rescatterings wash out the original emission region
and thus distort this measurement.
 
Although femtoscopy has been used widely to study a
number of processes ($hh, e^+e^-, AA$~\cite{rew}) this
is not the case for the cumulative process. 
Hadroproduction data exist for carbon and heavier
nuclei~\cite{BAS89,hadro}, but lepton-nucleus data
are scarce in any kinematical
domain~\cite{WA25_nd_pipi,ITEP90_eA}. In~\cite{WA25_nd_pipi}
the size of the pion emission region was studied in high energy
$\nu$D interactions.
In~\cite{ITEP90_eA} data 
on two-proton and two-pion correlations were obtained in
$e^{16}$O interactions at 5 GeV. The scattered electron was
not identified; the data correspond to a small $Q^2$ value and
to $\nu \sim 1.5$ GeV. The measured source size proved to be 
commensurate with the nuclear size and showed a tendency
to decrease with particle momenta.    

We present here our study of the correlation between
two detected protons with  small relative momenta in
$eA(^3$He, $^4$He, $^{12}$C, $^{56}$Fe)$\rightarrow e'ppX$  
reactions, for an incident electron energy of 4.46 GeV. 
The measurements were performed with the CEBAF Large
Acceptance Spectrometer (CLAS)~\cite{CLAS} in Hall B at the
Thomas Jefferson National Accelerator Facility.
The CLAS detector is a six-sector toroidal magnetic spectrometer.  The
magnetic  field is generated by iron-free superconducting coils.
The detection systems consist of drift chambers to  determine 
the  trajectories of charged  particles~\cite{DC}, 
scintillator counters for  time-of-flight measurements~\cite{SC}, 
Cherenkov counters to distinguish between electrons and
negative    pions~\cite{CC},    and electromagnetic   shower
calorimeter to identify electrons and neutrons~\cite{EC}.
The CLAS was triggered on scattered electrons detected in  the
calorimeter with an energy above 1 GeV. 

Run conditions are described in detail in Ref.~\cite{KIM}.
Events with transferred energy $\nu$ between 0.5 and 3.5 GeV and
transferred  4-momentum squared  $Q^2$ between 0.6 and 5 (GeV/c)$^2$ 
were selected for analysis.
Protons in the momentum  range from 0.3 to 1.0 GeV/c were 
selected for the analysis. The angle $\theta$ between the direction of  
the virtual photon  $\gamma$ and  the  direction of  the
detected proton was from  $0^{\circ}$ to  $115^{\circ}$. 
The analysis was performed for events with  at least
two  detected  protons. Misidentification of 
electrons or protons was negligible.

In this article we shall use the ``mixing'' procedure~\cite{K74}
for the correlation function(CF) calculations, {\it i.e.\ }
\begin{equation}
\label{Rdef}
R(q,p) =
\frac{N_r(q,p)}{N_m(q,p)},
\end{equation}
where $q = |\vec{q}|$, $p = |\vec{p}|$ and  $ \vec{p} =
 (\vec{p_1} + \vec{p_2})/2$;
 $N_r$  and $N_m$ are the  numbers of proton  pairs  from
the real events and those combined from protons taken from
different events, respectively.      
Secondary particles are boosted in the direction of
the virtual photon momentum.   
We  select the mixed-pair protons from 
events for which the magnitude of the momentum difference
of the scattered electrons
$| \vec{p_{e1}} - \vec{p_{e2}}|$ is less than $q_0$.
We studied the dependence of the $N_m$ distribution on 
the value of $q_0$, and found this dependence 
negligible for $q_0<$0.2 GeV/c; a cut $q_0<$0.2 GeV/c was
applied for the $N_m$ distributions.    
Pairs of tracks hitting a single
 scintillator were not included in  our analysis,
because they have ambiguous time-of-flight values.

\begin{figure}
 \centerline{
 \includegraphics[width=90mm]{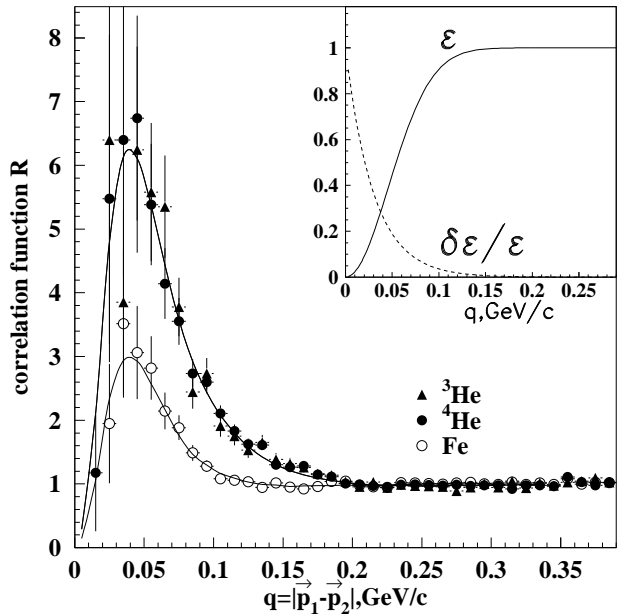}}
 \caption{
The two-proton correlation function $R$ for $^3$He, 
$^4$He and Fe nuclei. Curves are calculated for $r_{RMS}=$1.6 
fm (He) and $r_{\rm RMS} = $3.0 fm (Fe). The insertion shows 
the close-track efficiency $\varepsilon(q)$ and its uncertainty
$\delta\varepsilon/\varepsilon$.}
\label{corfunall}
\end{figure}

The ability to detect two tracks with small relative
momentum is limited because both  particles hit the same or
neighboring detector cells.  As a rule, the probability for 
losing at least one of the two tracks is higher for close tracks.
A  detailed study  of the close-track efficiency 
$\varepsilon(q)$  in CLAS has been  done  in Ref.~\cite{EFF}. It
 depends on track curvature and then,
for a fixed nominal magnetic field, on the proton momentum and 
emission angle in the laboratory system. The dependence 
of $\varepsilon(q)$ for the mean momentum and emission angle
is shown in the insert of  Fig.~\ref{corfunall}. 

Fig.~\ref{corfunall} shows $R(q)$ for the  $^3$He, $^4$He, 
and Fe data corrected for close-track efficiency $\varepsilon(q)$,
``long-range'' correlations(LRC), and momentum resolution. 
The data are averaged over proton momenta.
LRCs arise mainly from 
momentum conservation for the real events which 
is not a requirement for mixed pairs. 
It results in a smooth increase of $R$ with $q$, which 
reflects the fact that due to momentum conservation
the probability of particles emission in the same direction 
is smaller than that in the opposite direction. 
Empirically, LRC can be parameterized by
$R \propto \exp(b\cos\psi)$, in which $\psi$ is the 
angle between the two protons and a $b$ is a parameter~\cite{wide}.
The parameter $b$ for different $A$ and proton pair momenta,
was obtained from a fit to the data in which the region of
the effect at small $q<0.2$ GeV/c was cut out.
The correction for LRCs were made by 
introducing a weight $w = \exp(b\cos\psi)$ for mixed pairs
to reproduce LRC in the $N_m$ distribution.
The proton momentum resolution within the selected 
kinematic range is estimated to be $\delta p/p~ \sim~2~\%$.
Since $\delta p$ typically smaller than the width of the effects
under study measured correlation functions are only
slightly smeared out by momentum resolution. The
momentum resolution corrections were made by applying
the smearing procedure $n$ times to the measured CF and 
then by the extrapolation of the results to $n = -1$. 

Fig.~\ref{corfunall} also shows the theoretical
dependencies of $R(q)$~\cite{LL82} for $r_{\rm rms} = $ 1.6 
and 3.0 fm calculated within the model of independent 
one-particle sources taking into account quantum 
statistics and FSIs in the two-proton system.
The theoretical correlation function is then 
calculated as a square of the wave function 
(corresponding to the scattering problem)
averaged over the relative distances of the emitters 
in the pair rest frame. We assume a Gaussian distribution 
of the emission coordinates in the nucleus rest frame 
characterized by a dispersion $r_0^2 = r_{\rm rms}^2/3$. 
We neglect here the emission duration which enters
in the longitudinal component of the relative 
distance vector through the Lorentz transformation 
to the pair rest frame (the duration is thus effectively 
absorbed by the parameter  $r_{\rm rms}$). Both 
the correlation functions and the theoretical curves are 
normalized to unity for 0.17$<q<$0.35 GeV/c.
The theoretical approach~\cite{KOO77,LL82} predicts that 
the enhancement of $R$ at small $q$
is inversely related to the measured size parameter. 
The peak at $q \approx~ 0.04$ GeV/c results
mainly from the interplay between  
the attractive s-wave strong final-state interaction
and the Coulomb  repulsion.  
We compared the results of the calculation of the theoretical CF
for different proton-proton potentials~\cite{KOO77,LL82,CZECH86}: i.e. 
spherical  wave approximation (scattered wave $ \sim 1/r$),
simple  square well potential, 
Reid~\cite{REID}, and Tabakin~\cite{TABAKIN}.
For large $r_{\rm RMS}$ values, the correlation function is mainly
determined by the solution of the scattering problem outside the
range of the strong interaction potential, 
and is therefore independent
of the actual form of the potential, provided that it correctly
reproduces the scattering amplitudes~\cite{LL82,CZECH86}. 
Our results start to depend on the 
potential choice for  $r_{\rm RMS}<$2 fm.
At $r_{\rm RMS}<$2 fm the calculated curves for different 
potentials also look similar, but the best value of 
$r_{\rm RMS}$ depends on the version of the potential.  
In the present work final results for  $r_{\rm RMS}$
are presented for the realistic Reid potential, and 
the difference  between results calculated for the
potential with core (Reid)~\cite{REID} 
and without core (Tabakin)~\cite{TABAKIN}
($\approx 3\%$ in $r_{\rm RMS}$ for the He data) 
is taken as the theoretical uncertainty. 

The curves in Fig.~\ref{corfunall}
represent the best fit to the  data by the theoretical curves
(the difference between  $^3$He and $^4$He is negligible)
with $r_{\rm RMS}$ as a free parameter.
The fits in  Fig.~\ref{corfunall} are quite
reasonable. This is 
 an indication that the theoretical approach is applicable 
down to a measured size of the order 1.5 fm.
The dependencies of $R$ on $q$ for $^3$He and $^4$He
(and the best value for $r_{\rm RMS}$) are the
same within errors; the enhancement of $R$ at small $q$
for Fe is much smaller. This means $r_{\rm RMS}$
is larger for Fe than for He. The results for carbon 
(not shown in the figure) lie between He and Fe.

Experimental systematic errors on $r_{\rm rms}$ 
arise from the close-track
efficiency correction ($\approx 2\%$), 
the correction for ``long-range'' 
correlations ($\approx 2\%$),
and the correction for momentum resolution
($\approx 1\%$).
Non-identified $\Lambda$ particles that decay into $p\pi$ 
provide a potential background for the measured CF.
The cross section for $\Lambda$ production  is estimated
to be smaller than $ 1\%$ of the proton production
cross section in the corresponding kinematical region.
(Additional mass $\delta_m \sim m_K + m_{\Lambda} - m_p$ 
must be produced, and in a cumulative process the 
cross sections falls exponentially with $\delta_m$,
having a slope parameter on the order of the pion mass). 
Therefore,  the background from 
non-identified  $\Lambda\to p\pi$ decay is negligible. 
The total systematic experimental
errors on $r_{\rm RMS}$ is about $3\%$.
In the figure statistical and 
systematic errors have been added in quadrature.

\begin{figure} 
 \centerline{
 \includegraphics[width=90mm]{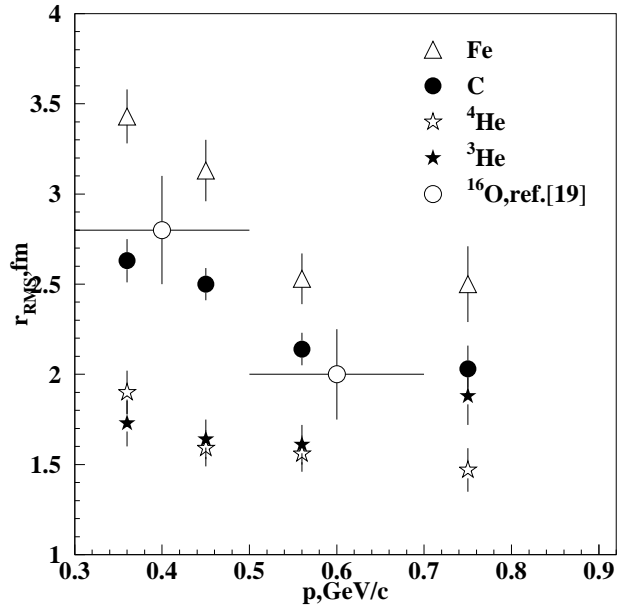}}
 \caption{The size parameter $r_{RMS}$ as a function of the mean
pair  momentum $p = |\vec{p}_1+\vec{p}_2|/2$. Data~\cite{ITEP90_eA},
which correspond to $e^{16}$O interactions at initial energy 5 GeV
and $Q^2<$ 0.1(GeV/c)$^2$, are shown for comparison.
}
\label{size-p}
\end{figure}

The  dependence of $r_{\rm RMS}$ on $p = |\vec{p}_1+\vec{p}_2|/2$ 
for different nuclei is shown in Fig. \ref{size-p}.  
The data are averaged over emission angles, statistical and 
systematic errors have been added in quadrature. 
For $^3$He the momentum dependence looks flat, while for 
carbon and iron it decreases with  
increasing pair momentum. Our results for carbon are in
good correspondence with the data~\cite{ITEP90_eA} for 
electron-oxygen interactions.
The values of $r_{\rm RMS}$ approach
the size of the nucleus for the 
lowest value of pair momentum,
which seems to be due to the rescattering of protons
in nuclear matter. The importance of rescattering  
decreases with proton momenta in the chosen momentum
range due at least in part to the decrease 
in the NN cross section. 

We estimate the size of the flucton $r_f$ under the assumption 
that both the primordial source size $r_f$ and its modification
due to rescattering processes contribute to the measured size. 
In the case of helium, the probability of rescattering 
is much smaller than in heavy nuclei.
The extracted $r_{\rm RMS}$ values in $^3$He and $^4$He are 
about the same, which is additional evidence that
rescattering does not affect the helium data within 
the errors ($\approx 0.1$ fm).
Therefore, $r_{\rm RMS}$ in helium
($\approx 1.6$ fm) is an upper estimate of $r_f$.

To take into account the possible influence of the 
rescattering process for helium,
we can extrapolate  the measured sizes as a function of A
to the minimum possible
target mass, where rescattering is not possible.
This will provide a lower estimate of $r_f$,
because rescattering can only increase the measured size.
The minimal target mass (in nucleon mass units) for the 
electro-production of protons(the so-called cumulative 
number $X_S$~\cite{VS}) 
is determined by the kinematics of the process  
$e+X_S \cdot m_p \rightarrow e'+p+m_c$
and
is given by:
\begin{equation}
X_S = \frac{\frac{Q^2}{2 \nu} + E_p - P_p \cos \Theta_{p \gamma} 
\sqrt{1+Q^2/ \nu^2}}
{(1-T_p/ \nu) m_p},
\label{XSTAV}
\end{equation}
in which $E_p$, $P_p$, $m_p$ and $T_p$ are the full energy, 
momentum, mass and kinetic energy of the proton, and 
$\theta_{p\gamma}$ is the angle 
between the proton momentum and the virtual photon 
momentum and $m_c$ is determined by conservation laws
for quantum numbers, baryon number in our case.
In the limit of large $\nu$, $X_S$ approaches the 
sum of the Bjorken variable 
$X_{Bj} = Q^2/2m_p\nu$ and the light cone 
variable $\alpha = (E_p -P_p \cos \theta_{p\gamma})/m_p$. 
For a di-proton (a proton pair at small relative momentum) 
the electro-production cumulative number 
is given by Eq.~\ref{XSTAV} in which 
$E_p$, $P_p$, $T_p$ and $\theta_{p\gamma}$
now refer to the pair.

Cumulative production is defined to occur when $X_S$ is 
larger then unity. Half of our proton pairs 
are produced with $X_S > 2$; the remaining events are still 
close to the kinematic boundary in the reaction
when the mass of the target is the two-nucleon mass.
An extrapolation of the measured sizes to $A \sim X_S$ 
provides the pair momentum average value $1.2 \pm 0.1$ fm,
where the error arise mainly from the dependence 
of the result on the extrapolation law.
The measured CF and then the extracted $r_{\rm RMS}$ could be affected 
by background from the decay of short-lived 
resonances like the $\Delta$. Since the proton velocity
in the $\Delta$ decay reference frame is small $v\sim 0.2$
and the lifetime is of the order $c\tau \approx 2$ fm, this background
could contribute $\sqrt{(r_{\rm RMS})^2+(v\tau)^2)} - r_{\rm RMS}$
to the measured size, which is less than 0.1 fm.
Given the maximum possible value
of this background the lower estimate for
the flucton size is 1 fm.
Therefore, we estimate the flucton size as 
$r_f = 1.3 \pm 0.3 $ fm, which is an
average of the 1 fm lower estimate and the measured
value for He of 1.6 above. 
It should also be noted that the 
flucton size estimate in~\cite{Bloha,Burov}
was indirect, rather imprecise, and  
based on the model for fitting inclusive data only. 
This work presents direct measurement 
of the flucton size.

In summary, the correlations  between   
protons produced  in  $eA$
interactions at 4.46 GeV have been investigated.
The data   clearly show a  narrow  
structure in the correlation function 
in  the  region  of small
relative  momenta ($q<$0.1 GeV/c) with 
a peak  at  $q\sim$0.04 GeV/c
which is in accordance with theoretical expectation. 
The helium data on  two-proton correlations at small
relative momentum have been obtained for the first time.
The measured size of the emission region 
 $r_{\rm RMS}$ depends on $A$ and the pair momentum.
Our estimate of the flucton size  
provides a value of $ r_f = 1.3 \pm 0.3 $ fm.

We would like to acknowledge the outstanding efforts
of the staff of the Accelerator and the Physics
Divisions at Jefferson Lab that made this experiment
possible. 
This work was supported in part by the Istituto Nazionale 
di Fisica Nucleare, the  French Centre National de la 
Recherche Scientifique, the French Commissariat \`{a} l'Energie 
Atomique, the U.S. Department of Energy, the National 
Science Foundation, Emmy Noether grant from the Deutsche 
Forschungs gemeinschaft, the Korean Science and Engineering 
Foundation, and the Grant Agency of the Czech Republic 
under contract 202/04/0793. 
The Southeastern Universities Research Association (SURA) 
operates the Thomas Jefferson National Accelerator 
Facility for the United States Department of Energy 
under contract DE-AC05-84ER40150.

\end{document}